\documentclass[english,preprint,tightenlines,eqsecnum,aps,prd,nofootinbib]{revtex4}
\usepackage[T1]{fontenc}
\usepackage[latin9]{inputenc}
\usepackage{babel}
\usepackage{amsmath}
\usepackage{amssymb}
\usepackage{graphicx}
\usepackage[unicode=true,pdfusetitle,
 bookmarks=true,bookmarksnumbered=false,bookmarksopen=false,
 breaklinks=false,pdfborder={0 0 1},backref=false,colorlinks=false]
 {hyperref}

\makeatletter
\@ifundefined{textcolor}{}
{%
 \definecolor{BLACK}{gray}{0}
 \definecolor{WHITE}{gray}{1}
 \definecolor{RED}{rgb}{1,0,0}
 \definecolor{GREEN}{rgb}{0,1,0}
 \definecolor{BLUE}{rgb}{0,0,1}
 \definecolor{CYAN}{cmyk}{1,0,0,0}
 \definecolor{MAGENTA}{cmyk}{0,1,0,0}
 \definecolor{YELLOW}{cmyk}{0,0,1,0}
}

\usepackage{braket}
\usepackage{caption}
\@ifundefined{showcaptionsetup}{}{%
 \PassOptionsToPackage{caption=false}{subfig}}
\usepackage{subfig}
\makeatother

\begin{document}

\title{Renormalized stress-energy tensor for stationary black holes}

\author{Adam Levi}

\address{Department of physics, Technion-Israel Institute of Technology,\\
Haifa 32000, Israel}
\begin{abstract}
We continue the presentation of the pragmatic mode-sum regularization
(PMR) method for computing the renormalized stress-energy tensor (RSET).
We show in detail how to employ the $t$-splitting variant of the
method, which was first presented for $\left\langle \phi^{2}\right\rangle _{ren}$,
to compute the RSET in a stationary, asymptotically-flat background.
This variant of the PMR method was recently used to compute the RSET
for an evaporating spinning black hole. As an example for regularization,
we demonstrate here the computation of the RSET for a minimally-coupled,
massless scalar field on Schwarzschild background in all three vacuum
states. We discuss future work and possible improvements of the regularization
schemes in the PMR method.
\end{abstract}
\maketitle

\section{Introduction}

Computation of the renormalized stress-energy tensor (RSET) $\left\langle T_{\alpha\beta}\right\rangle _{ren}$
is important for studying the effects of a quantum field on the metric.
For example, the field can be a minimally-coupled, massless scalar
field $\phi\left(x\right)$ satisfying the d'Alembertian equation
\begin{equation}
\square\phi=0.\label{eq: Intro: - Dalamertian equation}
\end{equation}
The back-reaction on the metric is described by the semiclassical
Einstein equation \footnote{Throughout this paper units $G=c=1$ are used, as well as $\left(-+++\right)$
signature.}
\begin{equation}
G_{\alpha\beta}=8\pi\left\langle T_{\alpha\beta}\right\rangle _{ren},\label{eq: Intro - semiclassical Einstein Eq}
\end{equation}
where $G_{\alpha\beta}$ is the Einstein tensor. The most notable
physical phenomenon described by Eqs. (\ref{eq: Intro: - Dalamertian equation})-(\ref{eq: Intro - semiclassical Einstein Eq})
is evaporation of a black hole (BH) due to Hawking radiation \cite{Hawking - Particle creation by black holes}.

The naive computation of the stress-energy tensor yields a divergent
mode-sum. In flat spacetime one can use the normal ordering operator
to regularize this divergence. Unfortunately, normal ordering is not
defined in curved spacetime. In 1965 DeWitt \cite{Dewitt - Dynamical theory of groups and fields}
used the point-splitting method (first introduced by Schwinger \cite{Schwinger})
to renormalize $\left\langle \phi^{2}\right\rangle $, and in 1976
Christensen \cite{Christensen 1976} adapted it for the computation
of the RSET. In the point-splitting method one can compute the renormalized
value of  a quantity which is quadratic in the field (or its derivatives)
at a point $x$, e.g. $\left\langle \phi^{2}\left(x\right)\right\rangle $,
by splitting the point $x$ into two points $x,x'$, and considering
a two-point function (TPF) of the form $\left\langle \phi\left(x\right)\phi\left(x'\right)\right\rangle $.
Then one subtracts from the TPF a known counter term and takes the
limit $x'\to x$.

Although the recipe for computing the RSET is known for four decades
it is still a great challenge to compute it generically, even for
a prescribed background. The difficulty is caused by the limit $x'\to x$.
This limit can be taken rather simply if the modes are known analytically,
however, in most cases, and in particular for BHs backgrounds, the
modes are known only numerically, which turns the limit into a very
difficult task.

To overcome this difficulty, Candelas, Howard, Anderson and others
\cite{Candelas - 1979 - phi2 Schwarzschild,Candelas =000026 Howard - 1984 - phi2 Schwrazschild,Howard - 1984 - Tab Schwarzschild,Anderson - 1990 - phi2 static spherically symmetric,Anderson - 1995 - Tab static spherically symmetric}
developed a method to implement point-splitting numerically. This
method relied on a forth order WKB, which is very hard to compute
in a Lorentzian background, so they have used Wick's rotation and
computed the RSET in the Euclidean sector. More recently, Taylor and
Breen introduced \cite{Peter taylor} a more sophisticated version
of the traditional method, which does not require a WKB approximation,
but is still using the Euclidean sector.

Unfortunately, the Euclidean sector does not exist for most backgrounds,
and so the aforementioned methods were developed for static, spherically-symmetric
backgrounds. This leaves out all possible dynamical backgrounds, which
are of great interest, as well as backgrounds of spinning BHs, such
as Kerr, which are stationary but not static.

Recently, Levi and Ori introduced \cite{Levi =000026 Ori - 2015 - t splitting regularization,Levi =000026 Ori - 2016 - theta splitting regularization,Levi =000026 Ori - 2016 - RSET}
the pragmatic mode-sum regularization (PMR) method. This method does
not rely on the Euclidean sector, and all computations are done directly
in the Lorentzian background. The PMR method requires the background
to admit a single symmetry (a Killing field), and takes different
variants depending on the symmetry exploited.

So far, two variants of PMR were introduced in detail for the computation
of $\left\langle \phi^{2}\right\rangle _{ren}$: the $t$-splitting
variant \cite{Levi =000026 Ori - 2015 - t splitting regularization},
for stationary backgrounds, and the Angular-splitting variant \cite{Levi =000026 Ori - 2016 - theta splitting regularization},
for spherically-symmetric backgrounds. A third variant, the $\varphi$-splitting
(also named the azimuthal-splitting), for axially-symmetric backgrounds
was introduced briefly \cite{Levi =000026 Ori - 2016 - RSET}.

Although the details of implementing the method to compute the RSET
were not given yet, results for the RSET obtained using the PMR method
were already presented: The RSET in Schwarzschild in the Unruh state
was displayed using all three variants \cite{Levi =000026 Ori - 2016 - RSET}.
In a more recent work, the $t$-splitting and $\varphi$-splitting
variants were used to compute the RSET in Kerr in the Unruh state
\cite{Levi Eilon Ori and DeMeent - 2016 - Kerr}.

In this paper we fill in the gaps for the $t$-splitting variant by
showing in detail how to utilize it to compute the RSET for an asymptotically-flat,
stationary background. For briefness we restrict our attention to
a minimally-coupled, massless scalar field. We then harness it to
compute the RSET in Schwarzschild in all three vacuum states.

Further details of how to implement PMR for the RSET in the angular-splitting
variant for spherically symmetric backgrounds, and in the $\varphi$-splitting
variant for axially-symmetric backgrounds will be given elsewhere
\cite{Preparation}.

This paper is organized as follows: Section \ref{sec: PS} presents
the basic point-splitting scheme for the RSET. Section \ref{sec: Regulzriztion}
presents the the implementation of $t$-splitting for the calculation
of the RSET and section \ref{sec: Computation in Schwarzschild} presents
the detailed computation of the RSET in Schwarzschild. In section
\ref{sec: Discussion} we discuss the PMR method in general as well
as future research.

\section{The point-splitting scheme\label{sec: PS}}

The computation of the RSET in the point-splitting scheme, as formulated
by Christensen \cite{Christensen 1976}, takes the form 
\begin{equation}
\left\langle T_{\mu\nu}\left(x\right)\right\rangle _{ren}=\lim_{x'\to x}\left[\left\langle T_{\mu\nu}\left(x,x'\right)\right\rangle -C_{\mu\nu}\left(x,x'\right)\right],\label{eq: PS: Tab basic limit}
\end{equation}
where we use $C_{\mu\nu}\left(x,x'\right)$ for Christensen's counter-term
tensor, which depends on the bi-scalar $\sigma\left(x,x'\right)$,
its covariant derivatives and on the metric. The bi-scalar $\sigma\left(x,x'\right)$
is defined to be half the geodesic distance squared, of the short
geodesic connecting $x$ and $x'$. The explicit expression for $C_{\mu\nu}\left(x,x'\right)$
is very long and can be found in Ref. \cite{Christensen 1976}. The
quantity $\left\langle T_{\mu\nu}\left(x,x'\right)\right\rangle $,
which we will name the two-point stress tensor (TPST) is analogous
to the TPF used in the regularization of $\left\langle \phi^{2}\right\rangle $.
If we denote the symmetric TPF (also known as the Hadamard function)
by 
\begin{equation}
G^{\left(1\right)}\equiv\left\langle \phi\left(x\right)\phi\left(x'\right)\right\rangle +\left\langle \phi\left(x'\right)\phi\left(x\right)\right\rangle =\left\langle \left[\phi\left(x\right)\phi\left(x'\right)\right]_{+}\right\rangle ,\label{eq: PS: symmetric TPF}
\end{equation}
where $\left[\right]_{+}$ denotes anti-commutation, then the TPST
explicit form for a minimally-coupled, massless scalar field is \footnote{For brevity we restrict our attention to a minimally-coupled, massless
scalar field. The expression for a general scalar field can be found
in Ref. \cite{Christensen 1976}, and for an electromagnetic field
in Ref \cite{Christensen 1978}.}
\begin{gather}
\left\langle T_{\mu\nu}\left(x,x'\right)\right\rangle =\frac{1}{4}\left(\bar{g}_{\mu}^{\beta'}G_{;\beta'\nu}^{\left(1\right)}+\bar{g}_{\nu}^{\beta'}G_{;\mu\beta'}^{\left(1\right)}\right)-\frac{1}{4}g_{\mu\nu}\bar{g}_{\sigma}^{\beta'}G_{;\beta'}^{\left(1\right);\sigma}.\label{eq: PS: TPST}
\end{gather}
Primed derivatives are taken at $x'$, e.g.
\[
G_{;\beta'\nu}^{\left(1\right)}=\left\langle \left[\phi_{;\nu}\left(x\right)\phi_{;\beta'}\left(x'\right)\right]_{+}\right\rangle ,
\]
 and $\bar{g}_{\mu}^{\beta'}$ is the bi-vector of parallel transport,
which transfers a vector at $x'$ to $x$, satisfying
\begin{equation}
\sigma_{;\mu}=-\bar{g}_{\mu}^{\beta'}\sigma_{;\beta'}.\label{eq: PS: bi-vector of parallel transport}
\end{equation}
For a concrete split it is straightforward to compute $\bar{g}_{\mu}^{\beta'}$
as a series in the separation, by solving the parallel transport equation
order by order. For further details on the bi-scalar $\sigma$, and
bi-vector of parallel transport $\bar{g}_{\mu}^{\beta'}$ we refer
to \cite{Christensen 1976}.

We now take the recipe by Christensen and rewrite it to obtain a form
which is more natural for concrete computations. Notice that the RHS
of Eq. (\ref{eq: PS: TPST}) has the form of a trace-reversed tensor.
It is convenient to use this structure to reverse the trace of the
equation. Denoting with a tilde a trace reversed tensor, Eq. (\ref{eq: PS: TPST})
takes the form
\[
\left\langle \tilde{T}_{\mu\nu}\left(x,x'\right)\right\rangle =\frac{1}{4}\left(\bar{g}_{\mu}^{\beta'}G_{;\beta'\nu}^{\left(1\right)}+\bar{g}_{\nu}^{\beta'}G_{;\mu\beta'}^{\left(1\right)}\right).
\]
Substituting the Hadamard function
\[
\left\langle \tilde{T}_{\mu\nu}\left(x,x'\right)\right\rangle =\frac{1}{4}\left(\delta_{\mu}^{\alpha}\bar{g}_{\nu}^{\beta'}+\delta_{\nu}^{\alpha}\bar{g}_{\mu}^{\beta'}\right)\left\langle \left[\phi_{,\alpha}\left(x\right)\phi_{,\beta'}\left(x'\right)\right]_{+}\right\rangle .
\]
Inserting it to the trace reversed version of Eq. (\ref{eq: PS: Tab basic limit})
one obtains
\[
\left\langle \tilde{T}{}_{\mu\nu}\left(x\right)\right\rangle _{ren}=\lim_{x'\to x}\left[\frac{1}{4}\left(\delta_{\mu}^{\alpha}\bar{g}_{\nu}^{\beta'}+\delta_{\nu}^{\alpha}\bar{g}_{\mu}^{\beta'}\right)\left\langle \left[\phi_{,\alpha}\left(x\right)\phi_{,\beta'}\left(x'\right)\right]_{+}\right\rangle -\tilde{C}_{\mu\nu}\left(x,x'\right)\right],
\]
which can also be written in the form
\[
\left\langle \tilde{T}{}_{\mu\nu}\left(x\right)\right\rangle _{ren}=\lim_{x'\to x}\left(\delta_{\mu}^{\alpha}\bar{g}_{\nu}^{\beta'}+\delta_{\nu}^{\alpha}\bar{g}_{\mu}^{\beta'}\right)\left[\frac{1}{4}\left\langle \left[\phi_{,\alpha}\left(x\right)\phi_{,\beta'}\left(x'\right)\right]_{+}\right\rangle -\frac{1}{2}\delta_{\alpha}^{\sigma}\left(\bar{g}^{-1}\right)_{\beta'}^{\rho}\tilde{C}_{\sigma\rho}\left(x,x'\right)\right].
\]
The limit of the parentheses is finite, as well as the limit of the
brackets. So we can take the limit of the parentheses, which is trivial,
since the limit of the bi-vector $\bar{g}_{\mu}^{\beta'}$ is the
unit matrix $\delta_{\mu}^{\beta'}$ (Ref. \cite{Christensen 1976}).
Bringing the parentheses back into the brackets yields
\begin{multline*}
\left\langle \tilde{T}{}_{\mu\nu}\left(x\right)\right\rangle _{ren}=\lim_{x'\to x}\left|\frac{1}{4}\left\langle \left[\phi_{,\mu}\left(x\right)\phi_{,\nu}\left(x'\right)\right]_{+}\right\rangle +\frac{1}{4}\left\langle \left[\phi_{,\nu}\left(x\right)\phi_{,\mu}\left(x'\right)\right]_{+}\right\rangle \right.\\
\left.-\frac{1}{2}\left(\left(\bar{g}^{-1}\right)_{\nu}^{\rho}\tilde{C}_{\mu\rho}\left(x,x'\right)+\left(\bar{g}^{-1}\right)_{\mu}^{\rho}\tilde{C}_{\nu\rho}\left(x,x'\right)\right)\right].
\end{multline*}
Note that here we deviated from the usual covariant bi-tensor notation,
which is constructed to allow two different coordinate systems at
$x$ and $x'$. We use a single coordinate system, and take derivative
according to the coordinate system of point $x$ on the field at point
$x'$. This approach is less systematic, but proves to be very useful
for computations.

For convenience we define a new quantity, which is a symmetrized version
of the Christensen tensor $\tilde{C}_{\sigma\rho}\left(x,x'\right)$
multiplied by the bi-vector $\left(\bar{g}^{-1}\right)_{\mu}^{\rho}$.
Henceforth we will call this quantity the counter-term
\begin{equation}
\tilde{L}_{\mu\nu}\left(x,x'\right)\equiv\frac{1}{2}\left(\left(\bar{g}^{-1}\right)_{\nu}^{\rho}\tilde{C}_{\mu\rho}\left(x,x'\right)+\left(\bar{g}^{-1}\right)_{\mu}^{\rho}\tilde{C}_{\nu\rho}\left(x,x'\right)\right),\label{eq: PS: Define L}
\end{equation}
and we conclude that
\begin{equation}
\left\langle \tilde{T}{}_{\mu\nu}\left(x\right)\right\rangle _{ren}=\lim_{x'\to x}\left[\frac{1}{4}\left\langle \left[\phi_{,\mu}\left(x\right)\phi_{,\nu}\left(x'\right)\right]_{+}\right\rangle +\frac{1}{4}\left\langle \left[\phi_{,\nu}\left(x\right)\phi_{,\mu}\left(x'\right)\right]_{+}\right\rangle -\tilde{L}_{\mu\nu}\left(x,x'\right)\right].\label{eq: PS: Tab limit with L}
\end{equation}

One example that illustrates how $\tilde{L}_{\mu\nu}$ captures better
the essence of the singularity is the trace of the RSET for a minimally-coupled,
massless scalar field in a Ricci-flat solution. The trace of Christensen's
tensor $C_{\sigma}^{\,\,\sigma}\left(x,x'\right)$ in such a case
has no singular piece. Nevertheless, if one tries to naively (i.e.
with no split) compute the trace of the stress tensor one finds it
diverges. The trace $L_{\sigma}^{\,\,\sigma}\left(x,x'\right)$ however,
has a singular term proportional to $1/\sigma\left(x,x'\right)$,
which corresponds directly to the divergence of the mode-sum. 

\section{Regularization of the stress tensor\label{sec: Regulzriztion}}

Continuing the line taken in \cite{Levi =000026 Ori - 2015 - t splitting regularization}
we would like to build a scheme that will enable the computation of
the RSET using the mode-sum at the coincidence, based on the analytic
point-splitting recipe given by Christensen. We claim it is possible
given that the space-time admits some symmetry (a Killing field),
and for each symmetry the scheme takes a different form. Here we present
the $t$-splitting variant which is applicable for asymptotically-flat,
stationary backgrounds. We use the coordinates $t,\,r,\,\theta,\,\varphi$,
and the metric $g_{\alpha\beta}$ depends on $r,\,\theta,\,\varphi$
but not on $t$. In the limit $r\to\infty$ the metric takes the form
\[
ds^{2}=-dt^{2}+dr^{2}+r^{2}d\Omega^{2},
\]
where $d\Omega^{2}\equiv d\theta^{2}+\sin^{2}\theta d\varphi^{2}$. 

We shall first address the case of a background with a regular center,
and in Sec. \ref{subsec: Eternal-BHs} we will show the simple adjustments
for the case of an eternal BH. In a stationary background one can
decompose the modes in the form
\begin{equation}
\phi\left(x\right)=\int_{0}^{\infty}d\omega\sum_{l=0}^{\infty}\sum_{m=-l}^{l}\left(f_{\omega lm}\left(x\right)a_{\omega lm}+f_{\omega lm}^{*}\left(x\right)a_{\omega lm}^{\dagger}\right)\label{eq: Reg: phi decomposition}
\end{equation}
\begin{equation}
f_{\omega lm}\left(x\right)=e^{-i\omega t}Y_{lm}\left(\theta,\varphi\right)\bar{\psi}_{\omega lm}\left(r,\theta,\varphi\right),\label{eq: Reg: modes f}
\end{equation}
where $a_{\omega lm}^{\dagger},\,a_{\omega lm}$ are the creation
and annihilation operators and $Y_{lm}\left(\theta,\varphi\right)$
are the spherical harmonics functions. In a non-eternal, asymptotically-flat
background one can choose the initial conditions for the modes $f_{\omega lm}$
at past null infinity to be the same as in flat space, which defines
a vacuum state $\left|0\right\rangle $ such that $a_{\omega lm}\left|0\right\rangle =0$
for all $\omega lm$. The modes evolve according to the d'Alembertian
equation
\[
\square f_{\omega lm}=0,
\]
which yields a differential equation for $\bar{\psi}_{\omega lm}\left(r,\theta,\varphi\right)$
that can in general be solved numerically. It is important to note
that, unless specifically stated otherwise, all the integrals over
$\omega$ in this paper are generalized integrals, as defined in Appendix
\ref{sec: Appendix - Generalized-integrals-oscillations}. 

It is very helpful to first examine the naive divergent calculation
one gets when trying to calculate the stress-tensor without splitting
the points
\begin{equation}
\left\langle \tilde{T}{}_{\mu\nu}\left(x\right)\right\rangle _{naive}=\frac{1}{2}\left\langle \left[\phi_{,\mu}\left(x\right)\phi_{,\nu}\left(x\right)\right]_{+}\right\rangle =\hbar\int_{0}^{\infty}d\omega\sum_{l=0}^{\infty}\sum_{m=-l}^{l}\Re\left\{ f_{\omega lm,\mu}\left(x\right)f_{\omega lm,\nu}^{*}\left(x\right)\right\} ,\label{eq: Reg: Naive mode-sum}
\end{equation}
where $\Re$ denoted the real part of a complex number. The sum over
$m$ is a finite sum, and due to asymptotic flatness the sum over
$l$ converges \footnote{If there is no asymptotic flatness, or alternatively if the computation
is done in the interior of an eternal BH, the sum over $l$ might
not converge. One can then employ an intermediate regularization technique
similar to the one used in the angular splitting \cite{Levi =000026 Ori - 2016 - theta splitting regularization}.}. So one can define
\begin{equation}
F_{\mu\nu}\left(\omega,x\right)\equiv\sum_{l=0}^{\infty}\sum_{m=-l}^{l}\Re\left\{ f_{\omega lm,\mu}\left(x\right)f_{\omega lm,\nu}^{*}\left(x\right)\right\} ,\label{eq: Reg: Define F}
\end{equation}
where the tensor $F_{\mu\nu}$ can be computed numerically. Inserting
it back to Eq. (\ref{eq: Reg: Naive mode-sum}) we get
\[
\left\langle \tilde{T}{}_{\mu\nu}\left(x\right)\right\rangle _{naive}=\hbar\int_{0}^{\infty}F_{\mu\nu}\left(\omega,x\right)d\omega,
\]
which is of course divergent for most of the non-vanishing components,
e.g. to leading order $F_{tt}\left(\omega,x\right)\propto\omega^{3}$.

Returning to the regularization scheme in Eq. (\ref{eq: PS: Tab limit with L}),
we choose to split the points in the direction of the symmetry. For
the stationary backgrounds considered in this paper we split in the
$t$ direction, so the points are
\[
x=(t,r,\theta,\varphi),\,x'=(t+\varepsilon,r,\theta,\varphi),
\]
and taking the limit $x'\to x$ corresponds to taking $\varepsilon\to0$.
For this type of split the modes satisfy
\[
f_{\omega lm}\left(x'\right)=f_{\omega lm}\left(x\right)e^{-i\omega\varepsilon}.
\]
Using this identity together with the definition of $F_{\mu\nu}\left(\omega,x\right)$
in Eq. (\ref{eq: Reg: Define F}) one can recast Eq. (\ref{eq: PS: Tab limit with L})
to get

\begin{equation}
\left\langle \tilde{T}{}_{\mu\nu}\left(x\right)\right\rangle _{ren}=\lim_{\varepsilon\to0}\left[\hbar\int_{0}^{\infty}F_{\mu\nu}\left(\omega,x\right)\cos\left(\omega\varepsilon\right)d\omega-\tilde{L}_{\mu\nu}\left(x,\varepsilon\right)\right],\label{eq: Reg: Tab - the transform}
\end{equation}
where the counter-term is now considered as a function of the point
$x$ and $\varepsilon$. If we expand it in $\varepsilon$ it takes
the general form

\begin{equation}
\frac{1}{\hbar}\tilde{L}_{\mu\nu}\left(\varepsilon\right)=a_{\mu\nu}\left(x\right)\varepsilon^{-4}+b_{\mu\nu}\left(x\right)\varepsilon^{-2}+c_{\mu\nu}\left(x\right)\varepsilon^{-1}+d_{\mu\nu}\left(x\right)\ln\left(\mu\varepsilon\right)+e_{\mu\nu}\left(x\right)+O\left(\varepsilon\right),\label{eq: Reg: L expension}
\end{equation}
where $a_{\mu\nu}\left(x\right),b_{\mu\nu}\left(x\right),c_{\mu\nu}\left(x\right),d_{\mu\nu}\left(x\right),e_{\mu\nu}\left(x\right)$
are local tensors that depend on the metric, and $\mu$ is the unknown
parameter that corresponds to the scale-ambiguity in the regularization
(see Ref. \cite{Christensen 1976}).

One can now decompose $\tilde{L}_{\mu\nu}\left(\varepsilon\right)$
to its Fourier components, using the following identities ($\gamma$
is Euler's constant) :

\begin{alignat*}{1}
\int_{0}^{\infty}\omega^{3}\cos\left(\omega\varepsilon\right)d\omega & =6\varepsilon^{-4}\\
\int_{0}^{\infty}\omega\cos\left(\omega\varepsilon\right)d\omega & =-\varepsilon^{-2}\\
\int_{0}^{\infty}\ln\left(\omega\right)\cos\left(\omega\varepsilon\right)d\omega & =-\frac{\pi}{2}\varepsilon^{-1}\\
\int_{0}^{\infty}\frac{1}{\omega+\mu e^{-\gamma}}\cos\left(\omega\varepsilon\right)d\omega & =-\ln\left(\mu\varepsilon\right)+O\left(\varepsilon\right).
\end{alignat*}
Inserting it back to Eq. (\ref{eq: Reg: Tab - the transform}) one
obtains
\begin{equation}
\left\langle \tilde{T}{}_{\mu\nu}\left(x\right)\right\rangle _{ren}=\hbar\lim_{\varepsilon\to0}\left[\int_{0}^{\infty}F_{\mu\nu}^{Reg}\left(\omega,x\right)\cos\left(\omega\varepsilon\right)d\omega\right]-\hbar e_{\mu\nu}\left(r\right),\label{eq: Reg: Tab - still with a limit}
\end{equation}
where
\begin{equation}
F_{\mu\nu}^{Reg}\left(\omega,x\right)\equiv F_{\mu\nu}\left(\omega,x\right)-F_{\mu\nu}^{Sing}\left(\omega,x\right)\label{eq: Reg: Freg}
\end{equation}
and
\begin{equation}
F_{\mu\nu}^{Sing}\left(\omega,x\right)\equiv\frac{1}{6}a_{\mu\nu}\omega^{3}-b_{\mu\nu}\omega-\frac{2}{\pi}c_{\mu\nu}\ln\left(\omega\right)-d_{\mu\nu}\frac{1}{\omega+\mu e^{-\gamma}}.\label{eq: Reg: Fsing}
\end{equation}
Interchanging the limit and integral in Eq. (\ref{eq: Reg: Tab - still with a limit})
we obtain
\begin{equation}
\left\langle \tilde{T}{}_{\mu\nu}\left(x\right)\right\rangle _{ren}=\hbar\int_{0}^{\infty}F_{\mu\nu}^{Reg}\left(\omega,x\right)d\omega-\hbar e_{\mu\nu}.\label{eq: Reg: Tab - final result}
\end{equation}
This interchanging is not trivial, and we address it in the next subsection.

We emphasis that the integral over $\omega$ in Eq. (\ref{eq: Reg: Tab - final result})
is a generalized integral, and converges as such. Namely, it might
contain oscillation at large $\omega$. To pragmatically compute the
generalized integral we use the method of self-cancellation which
is described in App. \ref{sec: Appendix - Generalized-integrals-oscillations}.
Once $\left\langle \tilde{T}{}_{\mu\nu}\left(x\right)\right\rangle _{ren}$
is computed one can trace-reverse it to obtain $\left\langle T_{\mu\nu}\left(x\right)\right\rangle _{ren}$.

\subsection{Blind spots}

The idea behind the interchanging of the limit with the integral in
Eq. (\ref{eq: Reg: Tab - still with a limit}) is that the decomposition
of the counter-term would contain all the information about the divergent
part of $F_{\mu\nu}\left(\omega,x\right)$, and so one can simply
set $\varepsilon$ to zero in the integral. Unfortunately it is not
necessarily true. One can imagine a function $B\left(\omega\right)$,
for which the integral
\[
\int_{0}^{\infty}B\left(\omega\right)\cos\left(\omega\varepsilon\right)d\omega
\]
is identically zero, for any $\varepsilon\neq0$. Yet at the coincidence,
namely $\varepsilon=0$, the integral diverges. These types of functions
we name \textit{blind spots }\footnote{A similar concept of blind-spots was introduced in the Angular-splitting
\cite{Levi =000026 Ori - 2016 - theta splitting regularization}}\textit{.} It is quite simple to find not just one, but a family of
blind spots
\[
B_{n}\left(\omega\right)=\omega^{2n},\,\,\,n\in0,1,2,3...
\]
With this concept in mind, we return to Eq. (\ref{eq: Reg: Tab - still with a limit})
and see that there might be a singular part in $F_{\mu\nu}\left(\omega,x\right)$
that one can not learn about from the counter-term. So the integral
over $F_{\mu\nu}^{Reg}\left(\omega,x\right)$ does not necessarily
have to converge, as $F_{\mu\nu}^{Reg}\left(\omega,x\right)$ might
contain blind spots. This is quite manageable, as one can classify
the problematic blind spot and simply remove it by using techniques
similar to self cancellation, discussed in App. \ref{sec: Appendix - Generalized-integrals-oscillations}. 

If we assume that the only blind spots are of the form $B_{n}\left(\omega\right)$
(although we do not prove it), then we expect only $\omega^{2}$ or
$\omega^{0}$ to appear, as we expect the most singular part in $F_{\mu\nu}\left(\omega,x\right)$
to correspond to the most singular term in the counter-term, which
is proportional to $\omega^{3}$. In reality, we did not come across
any blind spot in the $t$-splitting, so our best guess is that this
blind spots do not exist in $F_{\mu\nu}\left(\omega,x\right)$. 

Note that the family of blind spots we presented here was not a blind
spot in the calculation for $\left\langle \phi^{2}\right\rangle _{ren}$
in \cite{Levi =000026 Ori - 2015 - t splitting regularization}. This
is due to the fact that the counter-term used there was developed
to an arbitrary split, and in the case of the RSET the Christensen
tensor was calculated for a symmetric split only. One can also see
it by noticing that for $\left\langle \phi^{2}\right\rangle _{ren}$
the decomposition was carried out using $e^{i\omega\varepsilon}$
and here it is done using $\cos\left(\omega\varepsilon\right)$. Taking
$\omega^{0}$, which is an example for a blind spot in the case of
the RSET, it is obvious that
\[
\int_{0}^{\infty}d\omega\omega^{0}e^{i\omega\varepsilon}=\frac{i}{\varepsilon}\neq0.
\]
So it is clearly not a blind spot in the calculation of $\left\langle \phi^{2}\right\rangle _{ren}$.

\subsection{Eternal BHs \label{subsec: Eternal-BHs}}

In the case of an eternal BH, one has to introduce a second set of
creation and annihilation operators in the decomposition of the field,
which defines the boundary condition on the past horizon. In this
case the field decomposition (analogous to Eq. (\ref{eq: Reg: phi decomposition}))
takes the from
\begin{equation}
\phi\left(x\right)=\int_{0}^{\infty}d\omega\sum_{l=0}^{\infty}\sum_{m=-l}^{l}\left(f_{\omega lm}\left(x\right)a_{\omega lm}+f_{\omega lm}^{*}\left(x\right)a_{\omega lm}^{\dagger}+g_{\omega lm}\left(x\right)b_{\omega lm}+g_{\omega lm}^{*}\left(x\right)b_{\omega lm}^{\dagger}\right).\label{eq: Reg: phi decomposition eternal}
\end{equation}
Each of the sets of modes $f_{\omega lm}$ and $g_{\omega lm}$ defines
a vacuum state such that $a_{\omega lm}\left|0^{a}\right\rangle =b_{\omega lm}\left|0^{b}\right\rangle =0$,
and both sets satisfy the d'Alembertian equation. In addition both
$f_{\omega lm}$ and $g_{\omega lm}$ can be written in the form of
Eq. (\ref{eq: Reg: modes f})
\begin{gather}
f_{\omega lm}\left(x\right)=e^{-i\omega t}Y_{lm}\left(\theta,\varphi\right)\bar{\psi}_{\omega lm}^{f}\left(r,\theta,\varphi\right),\nonumber \\
g_{\omega lm}\left(x\right)=e^{-i\omega t}Y_{lm}\left(\theta,\varphi\right)\bar{\psi}_{\omega lm}^{g}\left(r,\theta,\varphi\right).\label{eq: Reg: modes f - Eternal}
\end{gather}

The regularization scheme presented above is unchanged for an eternal
BH, except for the definition of $F_{\mu\nu}\left(\omega,x\right)$
in Eq. (\ref{eq: Reg: Define F}), which now takes the form
\begin{equation}
F_{\mu\nu}\left(\omega,x\right)\equiv\sum_{l=0}^{\infty}\sum_{m=-l}^{l}\Re\left\{ f_{\omega lm,\mu}\left(x\right)f_{\omega lm,\nu}^{*}\left(x\right)+g_{\omega lm,\mu}\left(x\right)g_{\omega lm,\nu}^{*}\left(x\right)\right\} .\label{eq: Reg: Define F - Eternal}
\end{equation}

\section{Computation of the RSET in Schwarzschild \label{sec: Computation in Schwarzschild}}

In this section the method constructed in Sec. \ref{sec: Regulzriztion}
is utilized to compute the RSET in Schwarzschild spacetime, for a
minimally-coupled, massless scalar field. We show a detailed calculation
for one component of the RSET in the Boulware state at a specific
$r$ value, and then present the full results of the RSET, for various
$r$ values, in all three vacuum states. The Schwarzschild metric
is
\[
ds^{2}=-\left(1-\frac{2M}{r}\right)dt^{2}+\left(1-\frac{2M}{r}\right)^{-1}dr^{2}+r^{2}d\Omega^{2}.
\]

Following Sec. \ref{subsec: Eternal-BHs} (as Schwarzschild is an
eternal BH) we know that the decomposition of the field contains two
sets of modes $f_{\omega lm}$ and $g_{\omega lm}$. Due to the spherical
symmetry the expression for the modes given in Eq. (\ref{eq: Reg: modes f - Eternal})
can now be expressed more explicitly

\begin{gather}
f_{\omega lm}\left(x\right)=e^{-i\omega t}Y_{lm}\left(\theta,\varphi\right)\bar{\psi}_{\omega l}^{f}\left(r\right),\nonumber \\
g_{\omega lm}\left(x\right)=e^{-i\omega t}Y_{lm}\left(\theta,\varphi\right)\bar{\psi}_{\omega l}^{g}\left(r\right).\label{eq: Sch: modes}
\end{gather}
Consequently, the mode-sum in Eq. (\ref{eq: Reg: Define F - Eternal})
is simpler, because the sum over $m$ can be done analytically. 

It is useful to express $\bar{\psi}_{\omega l}^{f/g}\left(r\right)$
by another radial function with a different normalization
\[
\bar{\psi}_{\omega l}^{f/g}\left(r\right)=\frac{\psi_{\omega l}^{f/g}\left(r\right)}{\sqrt{4\pi\omega}r}.
\]
 And $\psi_{\omega l}^{f}\left(r\right)$ and $\psi_{\omega l}^{g}\left(r\right)$
satisfy the radial equation
\begin{equation}
\frac{d^{2}\psi_{\omega l}\left(r\right)}{dr_{*}^{2}}=-\left[\omega^{2}-V_{l}\left(r\right)\right]\psi_{\omega l}\left(r\right),\label{eq: Sch: radial eq}
\end{equation}
where the effective potential $V_{l}\left(r\right)$ is given by
\begin{equation}
V_{l}\left(r\right)=\left(1-\frac{2M}{r}\right)\left[\frac{l\left(l+1\right)}{r^{2}}+\frac{2M}{r^{3}}\right],\label{eq: Sch: potential}
\end{equation}
 and $r_{*}$ is the usual tortoise coordinate
\[
r_{*}=r+2M\ln\left|\frac{r}{2M}-1\right|.
\]

The general solution of the radial equation (\ref{eq: Sch: radial eq})
is spanned by two sets of basis solutions

\begin{gather}
\psi_{\omega l}^{in}\left(r\right)=\left\{ \begin{array}[t]{cc}
\tau_{\omega l}^{in}\,e^{-i\omega r_{*}}, & r_{*}\to-\infty\\
e^{-i\omega r_{*}}+\rho_{\omega l}^{in}\,e^{i\omega r_{*}},\,\,\,\,\,\, & r_{*}\to\infty
\end{array}\right.\nonumber \\
\psi_{\omega l}^{up}\left(r\right)=\left\{ \begin{array}[t]{cc}
e^{i\omega r_{*}}+\rho_{\omega l}^{up}\,e^{-i\omega r_{*}},\,\,\,\,\,\, & r_{*}\to-\infty\\
\tau_{\omega l}^{up}\,e^{i\omega r_{*}}, & r_{*}\to\infty
\end{array}\right.\label{eq: Schwarzschild calc - Basic solutions bounadry conditions}
\end{gather}
where $\tau_{\omega l},\rho_{\omega l}$ represent the transmission
and reflection amplitudes. The vacuum state associated with the solutions
$\psi_{\omega l}^{in}\left(r\right),\,\psi_{\omega l}^{up}\left(r\right)$
is the Boulware vacuum state, and its modes are usually denoted as
$f_{\omega lm}^{in}$ and $f_{\omega lm}^{up}$ instead of $f_{\omega lm}$
and $g_{\omega lm}$. Here we first consider the Boulware state, and
the computations of the other two vacuum states is presented in Sec.
\ref{subsec: Unruh-and-Hartle-Hawking}. Once $\psi_{\omega l}^{in}\left(r\right)$
and $\psi_{\omega l}^{up}\left(r\right)$ are calculated numerically,
the quantity $F_{\mu\nu}\left(\omega,x\right)$ can be computed according
to Eq. (\ref{eq: Reg: Define F - Eternal}). In the Boulware state
we can write specifically
\[
F_{\mu\nu}\left(\omega,x\right)\equiv\sum_{l=0}^{\infty}\sum_{m=-l}^{l}\Re\left\{ f_{\omega lm,\mu}^{in}\left(x\right)f_{\omega lm,\nu}^{in*}\left(x\right)+f_{\omega lm,\mu}^{up}\left(x\right)f_{\omega lm,\nu}^{up*}\left(x\right)\right\} .
\]

For the Schwarzschild metric, the constant tensors that compose $\tilde{L}_{\mu\nu}\left(x,r\right)$
are 
\[
a_{\mu}^{\,\nu}\left(r\right)=\frac{1}{2\pi^{2}\left(1-2M/r\right)^{2}}Diag\left\{ -3,\,1,\,1,\,1\right\} ,
\]
\[
b_{\mu}^{\,\nu}\left(r\right)=\frac{M}{12\pi^{2}r^{3}\left(1-2M/r\right)^{2}}Diag\left\{ 0,\,-2,\,1-\frac{3M}{r},\,1-\frac{3M}{r}\right\} ,
\]

\[
c_{\mu}^{\,\nu}\left(r\right)=0,\,d_{\mu}^{\,\nu}\left(r\right)=0,
\]

\begin{multline*}
e_{\mu}^{\,\nu}\left(r\right)=\frac{M^{2}}{1440\pi^{2}r^{6}\left(1-2M/r\right)^{2}}Diag\left\{ \frac{69M^{2}-60Mr+12r^{2}}{r^{2}},\,\frac{393M^{2}-456rM+140r^{2}}{r^{2}},\right.\\
\left.\frac{87M^{2}-90rM+26r^{2}}{r^{2}},\,\frac{87M^{2}-90rM+26r^{2}}{r^{2}}\right\} ,
\end{multline*}
where $Diag$ is a shorthand for a diagonal matrix. Using these tensors
one can easily compute $F_{\mu\nu}^{Sing}\left(\omega,r\right)$ according
to Eq. (\ref{eq: Reg: Fsing}).

We have numerically computed the tensor $F_{\mu\nu}\left(\omega,r\right)$
by solving the radial equation (\ref{eq: Sch: radial eq}) using MATHEMATICA
and computing $\psi_{\omega l}^{in}\left(r\right)$ and $\psi_{\omega l}^{up}\left(r\right)$
for $\omega$ between $0$ and $3.5$ with a uniform step of $d\omega=1/300$.
For each $\omega$ the sum over $m$ was done analytically, and the
sum over $l$ was computed by sequentially computing $l$'s until
the sum converged to an accuracy of one part in $10^{12}$. Note that
in all the numerical values and plots we are using units $M=1$, in
addition to $G=c=1$. 

\begin{figure}
\subfloat[The solid line is the numerically calculated $F_{rr}\left(\omega,r=6\right)$;
it is proportional to $\omega^{3}$ for large $\omega$ and clearly
divergent. The dashed line is the analytically computed singular part,
$F_{rr}^{Sing}\left(\omega,r=6\right)$, which captures the non-oscillatory
divergent part of $F_{rr}\left(\omega,r=6\right)$.\label{fig:1a}]{\begin{centering}
\includegraphics[bb=30bp 180bp 560bp 620bp,clip,scale=0.4]{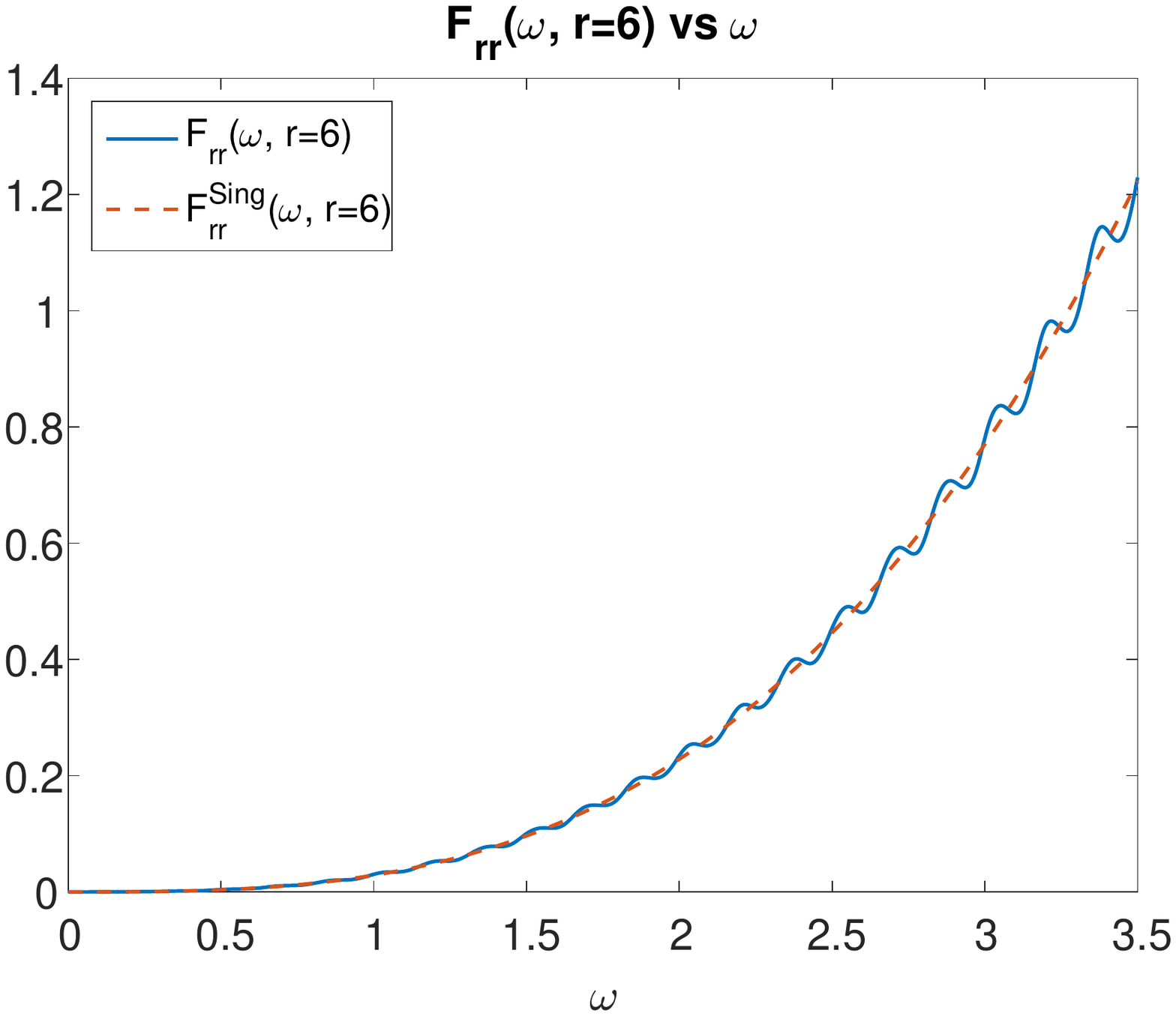}
\par\end{centering}
}\enskip{}\hfill{}\enskip{}\subfloat[The result of subtracting the singular part $F_{rr}^{Sing}\left(\omega,r=6\right)$
from $F_{rr}\left(\omega,r=6\right)$, which is the definition of
$F_{rr}^{Reg}\left(\omega,r=6\right)$ in Eq. (\ref{eq: Reg: Freg}).
It is dominated by oscillations with an amplitude proportional to
$\omega^{5/2}$. \label{fig:1b}]{\begin{centering}
\includegraphics[bb=30bp 180bp 560bp 620bp,clip,scale=0.4]{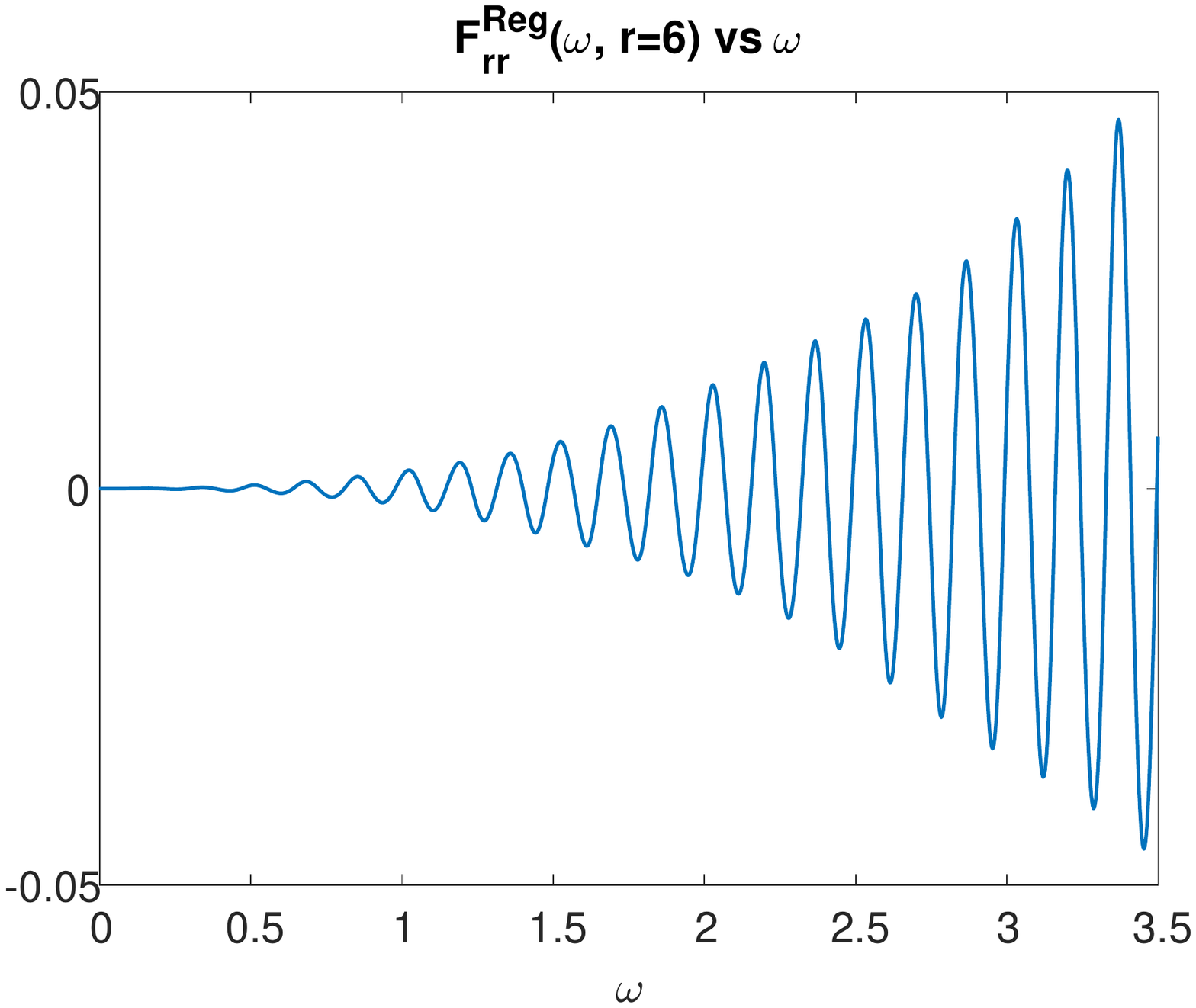}
\par\end{centering}
}

\caption{}
\end{figure}

The regularization was carried out according to Eq. (\ref{eq: Reg: Tab - final result}),
for the singular components of $\left\langle \tilde{T}{}_{\mu\nu}\left(r\right)\right\rangle _{ren}$
at various $r$ values. To demonstrate the scheme, regularization
of the $rr$ component is presented here for $r=6$ . Figure \ref{fig:1a}
exhibits the numerically calculated $F_{rr}\left(\omega,r=6\right)$,
together with the analytically computed $F_{rr}^{Sing}\left(\omega,r\right)$
from Eq. (\ref{eq: Reg: Fsing}). It is obvious that the integral
over $F_{rr}\left(\omega,r=6\right)$ will diverge. By subtracting
from it $F_{rr}^{Sing}\left(\omega,r\right)$, according to Eq. (\ref{eq: Reg: Freg}),
one obtains $F_{rr}^{Reg}\left(\omega,r\right)$ which is displayed
in Fig. \ref{fig:1b}.\\

\begin{figure}
\subfloat[The integral function $H_{rr}\left(\omega,r=6\right)$ vs $\omega$.
The amplitude of the oscillations is proportional to $\omega^{5/2}$.\label{fig:2a}]{\begin{centering}
\includegraphics[bb=30bp 180bp 560bp 620bp,clip,scale=0.4]{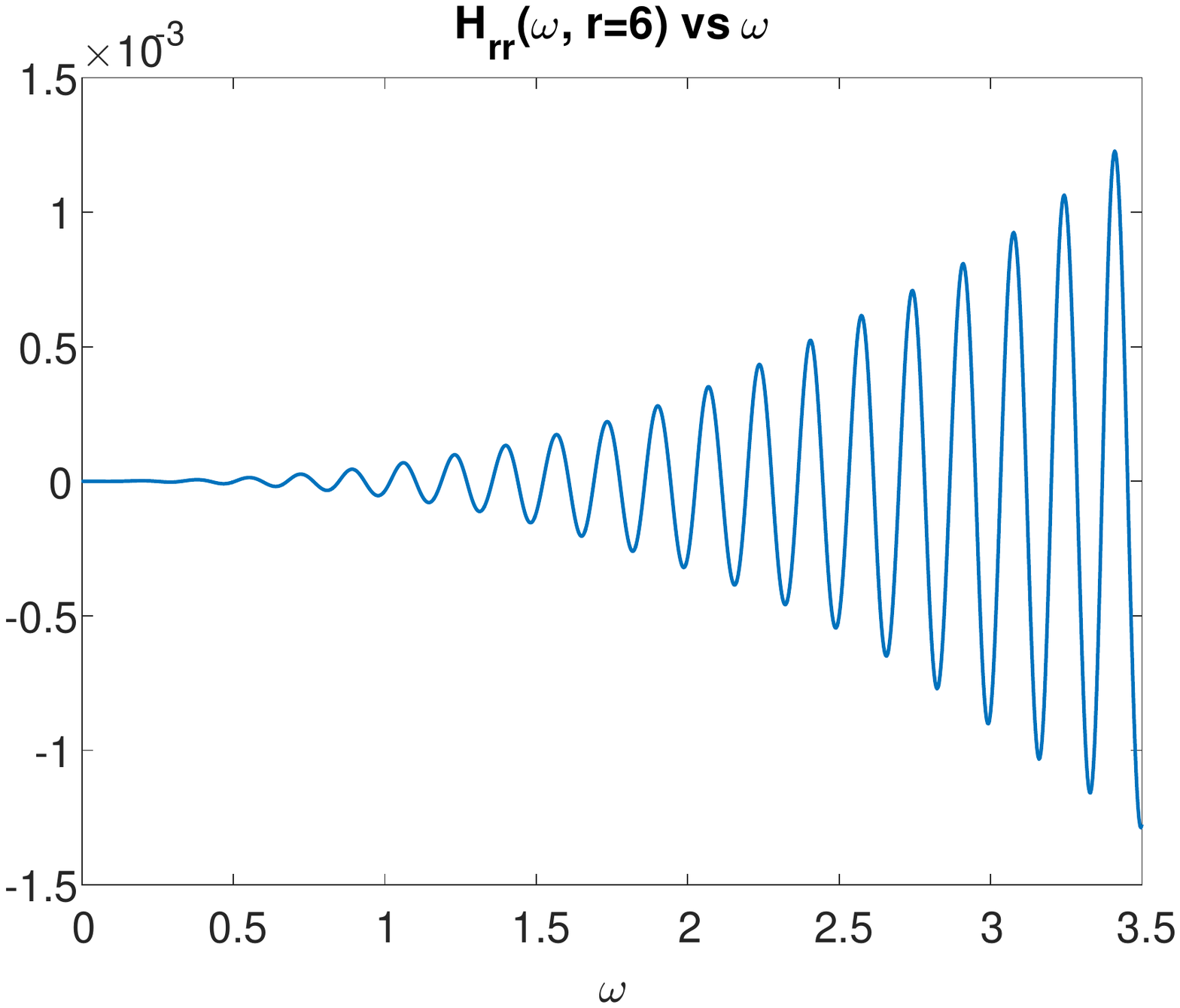}
\par\end{centering}
}\enskip{}\hfill{}\enskip{}\subfloat[$H_{rr}^{*}\left(\omega,r=6\right)$ vs $\omega$, obtained by applying
self-cancellation on $H_{rr}\left(\omega,r=6\right)$. Notice the
quick convergence and also the numerical deviations at $\omega\gtrsim1.5$.
The red cross denotes the point chosen by an algorithm we designed
to pick the optimal $\omega$ value to estimate the $\omega\to\infty$
limit. \label{fig:2b}]{\begin{centering}
\includegraphics[bb=30bp 180bp 560bp 620bp,clip,scale=0.4]{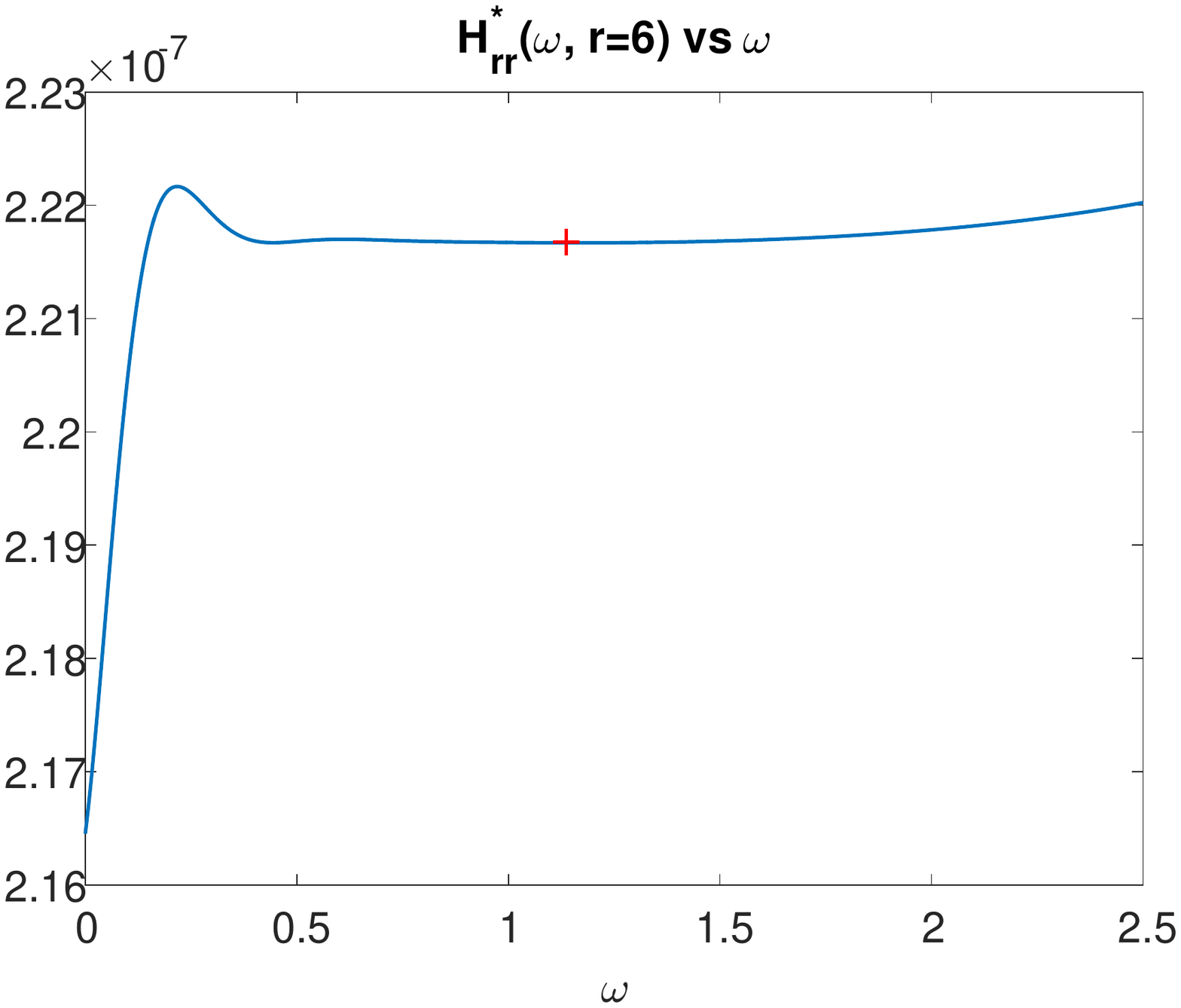}
\par\end{centering}
}

\caption{}
\end{figure}

The generalized integral over $F_{rr}^{Reg}\left(\omega,r=6\right)$
now converges, but it is clear that the regular integral does not.
To compute the generalized integral we employ the self-cancellation
technique that was presented in \cite{Levi =000026 Ori - 2015 - t splitting regularization},
and is also discussed in App. \ref{sec: Appendix - Generalized-integrals-oscillations}.
We first compute the regular integral function
\[
H_{rr}\left(\omega,r\right)=\int_{0}^{\omega}F_{rr}^{Reg}\left(\omega',r\right)d\omega',
\]
which can be observed in Fig. \ref{fig:2a} for $r=6$. It is now
possible to remove the oscillations by applying the self-cancellation
operator $T_{*}$, following Eq. (\ref{eq: App: Self-cancellation})
one can define
\[
H_{rr}^{*}\left(\omega,r\right)\equiv T_{*}\left[H_{rr}\left(\omega,r\right)\right].
\]

According to the self-cancellation technique the generalized integral
over $F_{rr}^{Reg}\left(\omega,r\right)$ is equivalent to the limit
$\omega\to\infty$ of $H_{rr}^{*}\left(\omega,r\right)$. Figure \ref{fig:2b}
displays $H_{rr}^{*}\left(\omega,r=6\right)$ and it is clear that
it has a defined limit as $\omega\to\infty$, and that it converges
to this limit rapidly. Notice that we have started with a quantity
($F_{rr}\left(\omega,r=6\right)$) with a magnitude of about unit
value, and after subtractions and self-cancellations obtained a result
($H_{rr}^{*}\left(\omega,r=6\right)$) of the order of about $10^{-7}$.
This is a great demonstration of the many orders of accuracy lost
in the regularization process. This loss of accuracy turns the computation
of the RSET into a difficult task, and requires the initial modes
to be computed with a very high accuracy. The loss of accuracy is
the cause of the numerical deviations at large $\omega$ in $H_{rr}^{*}\left(\omega,r=6\right)$
which are visible in Fig. \ref{fig:2b}. To handle this inaccuracy
we have built an algorithm that picks the optimal $\omega$ for evaluating
the limit $\omega\to\infty$. The chosen value is marked in Fig. \ref{fig:2b}
by a cross.\\

\begin{figure}
\begin{centering}
\includegraphics[bb=30bp 180bp 560bp 620bp,clip,scale=0.4]{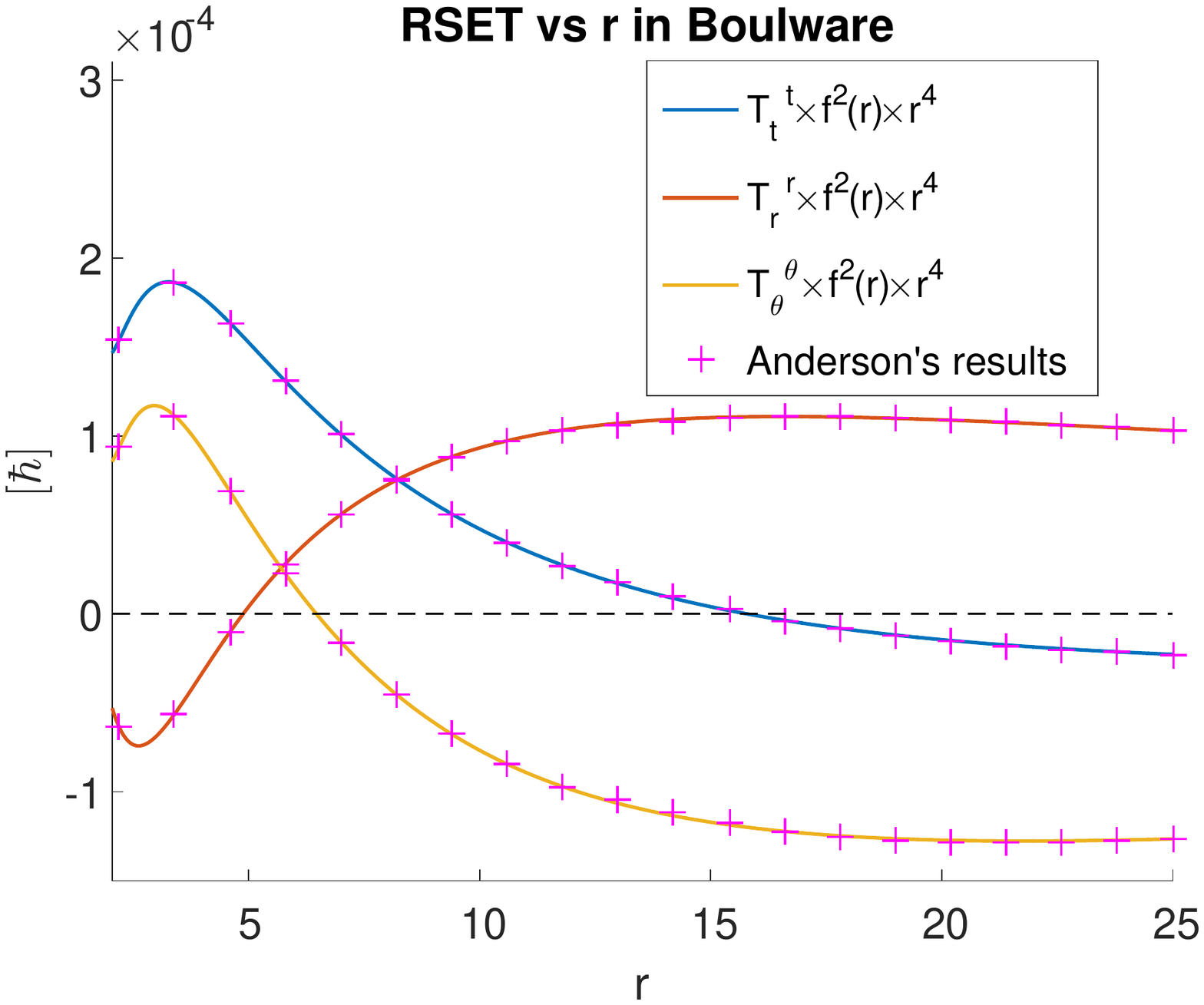}
\par\end{centering}
\caption{The solid curves represent the RSET in the Boulware state, calculated
using the $t$-splitting variant. The plus symbols are results by
Paul Anderson, obtained using the traditional regularization method
\cite{Anderson - 1995 - Tab static spherically symmetric}. The RSET
in the Boulware state is divergent at the horizon, and decays at infinity
like $r^{-4}$. To clarify the picture we multiplied the components
by factors of $f\left(r\right)\equiv1-2M/r$ and $r$.\label{fig:3}}
\end{figure}

Following Eq. \ref{eq: Reg: Tab - final result}, if we subtract from
the limit of $H_{rr}^{*}\left(\omega,r=6\right)$ the term $e_{rr}\left(r=6\right)$
we get $\left\langle \tilde{T}{}_{\mu\nu}\left(r=6\right)\right\rangle _{ren}$
(in units of $\hbar$). Repeating the scheme described above for different
$r$ values, and for all the non-trivial components of the RSET we
computed $\left\langle \tilde{T}{}_{\mu\nu}\left(r\right)\right\rangle _{ren}$.
Simply by trace-reversing this result we obtained $\left\langle T_{\mu\nu}\left(r\right)\right\rangle _{ren}$.
Figure \ref{fig:3} presents the results for the RSET, for all the
non-trivial components ($T_{t}^{\,t},T{}_{r}^{\,r},T{}_{\theta}^{\,\theta}=T{}_{\varphi}^{\,\varphi}$).
For reference we also plot results by Paul Anderson, obtained using
the traditional regularization method in the Euclidean sector \cite{Anderson - 1995 - Tab static spherically symmetric}.
We note that we also computed the RSET using the angular-splitting
and the agreement between the two is usually a few parts in $10^{3}$.
\\

\subsection{Unruh and Hartle-Hawking states\label{subsec: Unruh-and-Hartle-Hawking}}

Due to the fact that our method does not resort to the Euclidean sector,
the regularization process is exactly the same for all quantum states.
Thus in order to compute a different quantum state one only needs
to insert the solution for the modes in this particular state. In
the Schwarzschild example the two states that are physically interesting,
apart from the Boulware state, are the Unruh and Hartle-Hawking states.
Conveniently, in the Schwarzschild case one is not required to recalculate
the modes, and can build the the mode-sum in the Unruh and Hartle-Hawking
states using the modes of the Boulware state $f_{\omega lm}^{in}\left(x\right),\,f_{\omega lm}^{up}\left(x\right)$
\cite{Christensen and Fulling}. In our notation it can be written
as
\[
F_{\mu\nu}^{Unruh}\left(\omega,x\right)=\sum_{l=0}^{\infty}\sum_{m=-l}^{l}\Re\left\{ f_{\omega lm,\mu}^{in}\left(x\right)f_{\omega lm,\nu}^{in*}\left(x\right)+\coth\left(\frac{\pi\omega}{\kappa}\right)f_{\omega lm,\mu}^{up}\left(x\right)f_{\omega lm,\nu}^{up*}\left(x\right)\right\} .
\]

\[
F_{\mu\nu}^{H-H}\left(\omega,x\right)=\sum_{l=0}^{\infty}\sum_{m=-l}^{l}\coth\left(\frac{\pi\omega}{\kappa}\right)\Re\left\{ f_{\omega lm,\mu}^{in}\left(x\right)f_{\omega lm,\nu}^{in*}\left(x\right)+f_{\omega lm,\mu}^{up}\left(x\right)f_{\omega lm,\nu}^{up*}\left(x\right)\right\} .
\]

\begin{figure}
\begin{centering}
\includegraphics[bb=30bp 180bp 560bp 620bp,clip,scale=0.4]{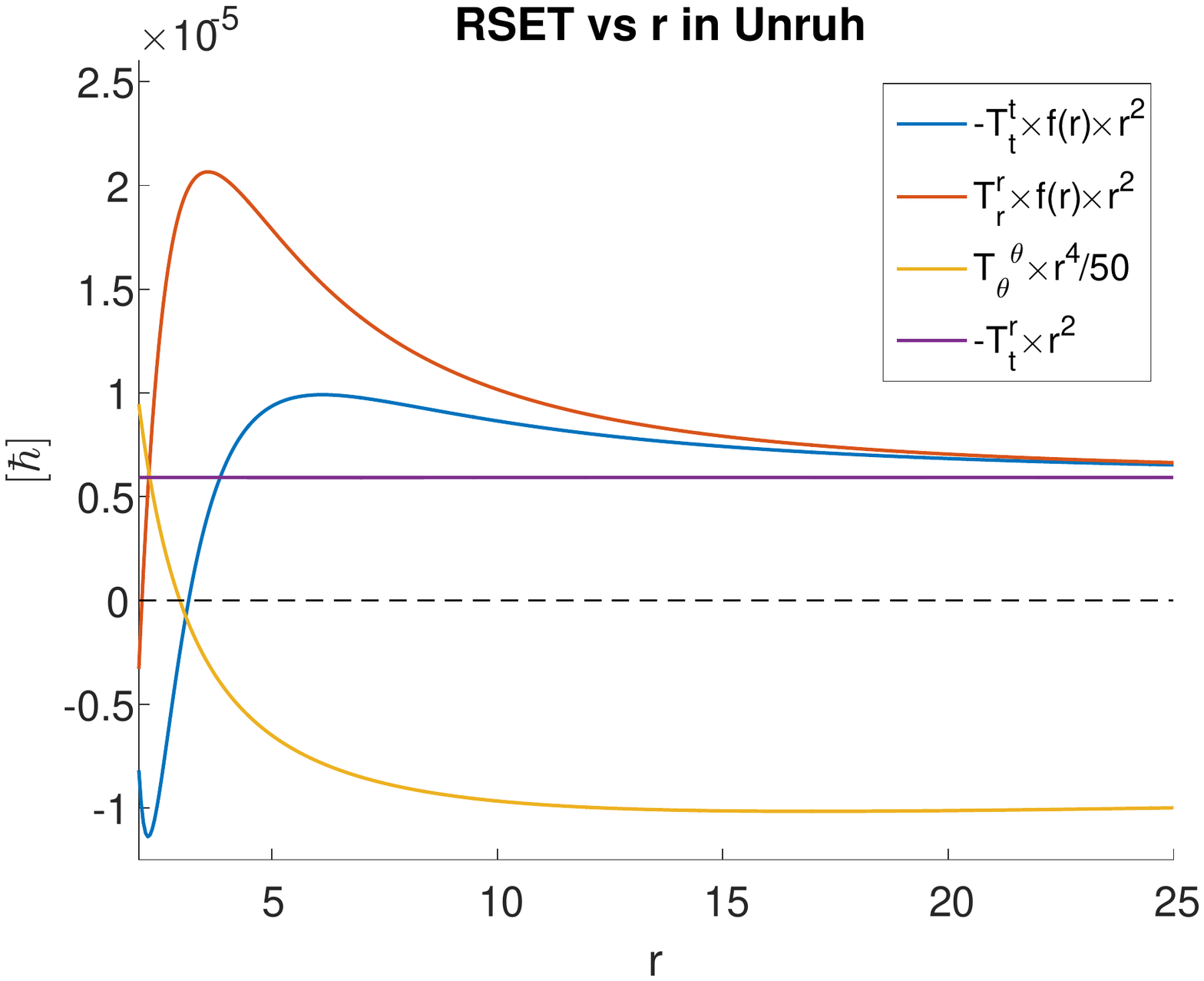}
\par\end{centering}
\caption{The RSET in the Unruh state. Here the RSET is finite on the horizon
and we multiply by $f\left(r\right)\equiv1-2M/r$ to treat the coordinate
singularity. Note that $-r^{2}T_{t}^{r}$ is a conserved quantity,
representing the outgoing Hawking radiation to infinity.\label{fig:4}}
\end{figure}

\begin{figure}
\begin{centering}
\includegraphics[bb=30bp 180bp 560bp 620bp,clip,scale=0.4]{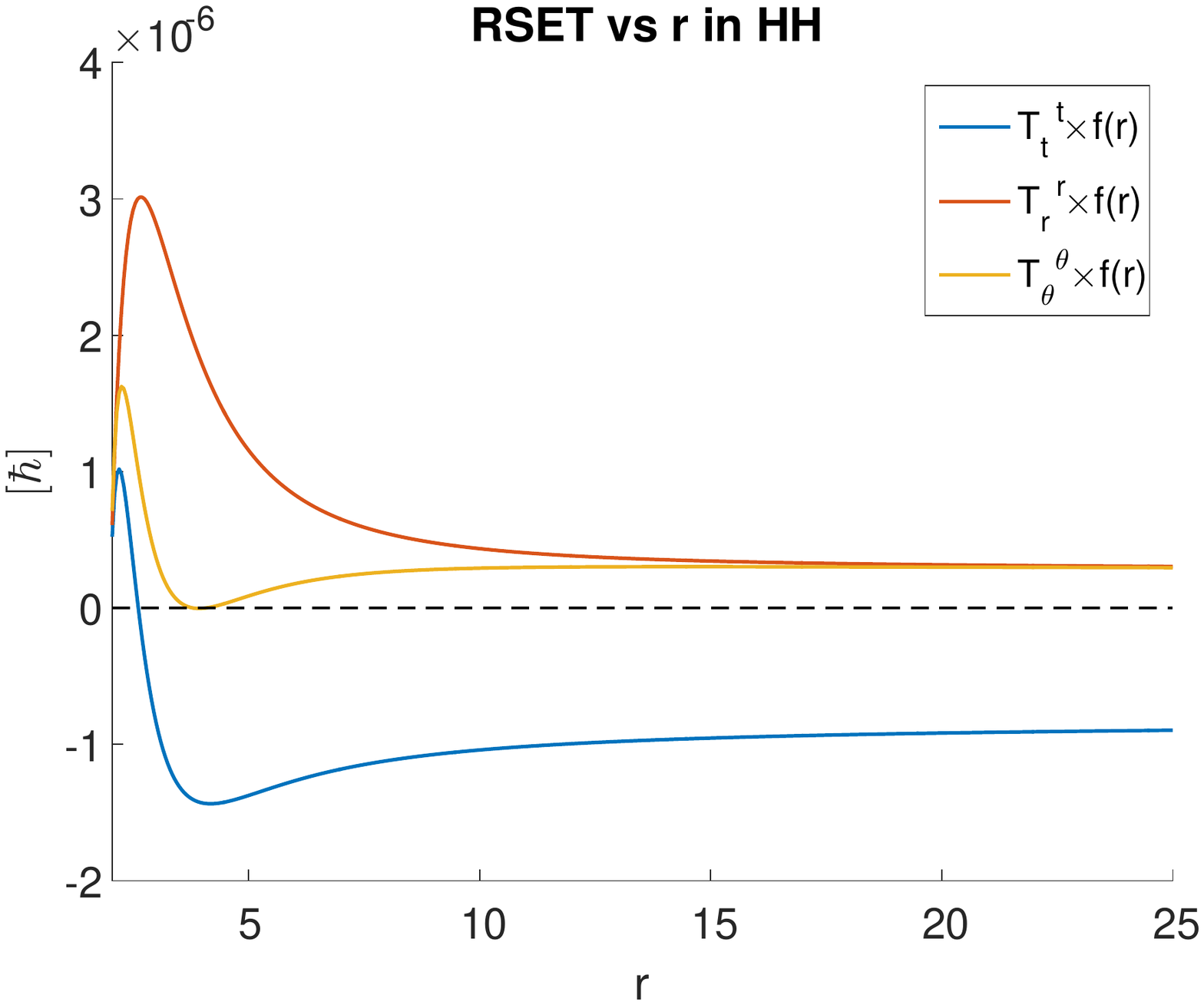}
\par\end{centering}
\caption{The RSET in the Hartle-Hawking state. Here the RSET is finite at the
horizon, and the multiplication by $f\left(r\right)\equiv1-2M/r$
is for scaling purposes only.\label{fig:5}}
\end{figure}

Using these relations we have calculated the RSET in the Unruh and
Hartle-Hawking states. Note that the regularization scheme remains
the same, because the counter-term is independent of the quantum state.
Figure \ref{fig:4} presents the results for the Unruh state and Fig.
\ref{fig:5} the results for the Hartle-Hawking state. In both states
we have crossed our results by computing the RSET in a different method,
the angular-splitting variant \cite{Levi =000026 Ori - 2016 - theta splitting regularization}.
In Unruh the two variants agree to about one part in $10^{3}$, and
in Hartle-Hawking the deviation is about one part in $10^{4}$. In
the Unruh state there is also a non-vanishing flux component $T_{t}^{r}$,
which represents the Hawking radiation. We computed the total luminosity
to be $L\equiv-4\pi r^{2}T_{t}^{r}\cong7.4388\cdot10^{-5}\hbar M^{-2}$,
in full agreement with the result by Elster \cite{Elster}. \\

\section{Discussion \label{sec: Discussion}}

We have presented the details of the RSET computation for stationary,
asymptotically-flat backgrounds, using the $t$-splitting variant
of the PMR method. Results were given for the Schwarzschild background
in all vacuum states for a minimally-coupled, massless scalar field.
We have also demonstrated that many orders of magnitude of accuracy
are lost in the computation, which requires the original modes to
be computed very accurately in order to obtain a good evaluation of
the RSET (see Sec. \ref{sec: Computation in Schwarzschild}).

Although the scheme was presented here for a minimally-coupled, massless
scalar field, it can also be implemented for other fields. For example,
one can use the same technique on Christensen's tensor for an electromagnetic
field, which can be found in \cite{Christensen 1978}. In addition
we have presented the method using Christensen's scheme and counter
terms, but one can also implement it to the Hadamard regularization
approach \cite{Wald - Hadamrad regularization}, where it can also
be extended to higher dimensions (similar to Ref. \cite{Peter taylor}).

An important issue regarding the $t$-splitting variant is to understand
where it is expected to break down. We do not go into deep investigation
of this point, and merely state that the scheme breaks down where
the norm of the Killing field vanishes. For example, in Schwarzschild
$t$-splitting does not work on the horizon. Furthermore, the scheme
does not break dichotomically, rather, it becomes less efficient on
approaching the locus where the norm vanishes. By less efficient we
mean that it requires a larger number of modes, namely a larger $l_{max}$
for the convergence of the sum, and a larger $\omega_{max}$ for the
convergence of the integral. For example, in Schwarzschild we used
$t$-splitting to approached the horizon up to $r=2.05M$.

Another important aspect of the $t$-splitting is the different techniques
for treating the oscillations. In App. \ref{sec: Appendix - Generalized-integrals-oscillations}
we discussed different techniques that we have explored and were not
mentioned in Ref. \cite{Levi =000026 Ori - 2015 - t splitting regularization}.
We believe that one can think of other techniques to implement the
generalized integral. As these techniques will become more efficient
the number of modes required for regularization will decrease and
computation will become easier. One such approach that we have not
tested yet is studying the oscillations using the \emph{global Hadamard
form. }Such global function was recently found for the Schwarzschild
case by Casals and Nolan \cite{Casals =000026 Nolan - Global Hadamrd function}.
We hope that future work will enable to study the global Hadamard
function for more generic backgrounds.

The $t$-splitting presented here was already used to compute $\left\langle \phi^{2}\right\rangle _{ren}$
and the RSET in the exterior region of Kerr \cite{Levi Eilon Ori and DeMeent - 2016 - Kerr},
yet we believe that there are many more implications for it. Possible
extensions of the current work in Kerr spacetime are to compute the
RSET inside the ergosphere and inside the horizon. In addition, the
method can be used to study other interesting stationery backgrounds
such as compact, rapidly spinning stars.

Other implementations of the PMR method, including the details of
how to compute the RSET using the angular-splitting variant, and the
details of the $\varphi$-splitting variant will be given elsewhere
\cite{Preparation}. Furthermore, we think that the different PMR
variants can be improved further in the future to create more efficient
schemes, without demanding more symmetries.

\section*{Acknowledgment}

I would like to thank Amos Ori, my Ph.D. supervisor, for all his guidance
and advice. I am also grateful to Paul Anderson for many interesting
discussions and for sharing with me his unpublished data.

\appendix

\section{Oscillations, generalized integrals, and self-cancellation \label{sec: Appendix - Generalized-integrals-oscillations}}

This appendix briefly summarizes the broad discussion in \cite{Levi =000026 Ori - 2015 - t splitting regularization}
regarding the oscillations in the mode sum. As explained there these
oscillations are caused by singularities in the TPF which are not
at the coincidence limit. Rather, these singularities correspond to
connecting null geodesics (CNGs) \cite{Kay Radzikowski and Wald - 1996},
and by computing the CNGs of the background metric one can compute
the wavelengths of the oscillations.

Analytically the oscillations are not a problem at all, as our integral
is a generalized integral. One way to define a generalized integral
over a function $h\left(\omega\right)$ is taking the limit of a Laplace
transform
\[
\lim_{\alpha\to0^{+}}\int_{0}^{\infty}h\left(\omega\right)e^{-\alpha\omega}d\omega.
\]
But this is very hard to perform numerically. To this purpose we have
introduced the self cancellation technique, which enables one to compute
the generalized integral for the case of a function with oscillations,
given that the wavelengths of the oscillations are known. For example,
if one wants to compute the generalized integral over $h\left(\omega\right)$,
which apart from a regularly convergent part contains an oscillatory
term of the type $\cos\left(\frac{2\pi}{\lambda}\omega\right)$, than
by first computing the standard integral function
\[
H\left(\omega\right)=\int_{0}^{\omega}h\left(\omega'\right)d\omega',
\]
one can apply the self-cancellation operator
\[
T_{\lambda}\left[H\left(\omega\right)\right]\equiv H\left(\omega\right)+H\left(\omega+\lambda/2\right)
\]
and the limit $\omega\to\infty$ of $T_{\lambda}\left[H\left(\omega\right)\right]$
will converge to the value of the generalized integral.

If the function $h\left(\omega\right)$ contains more than a single
wavelength, one can simply repeat this process for each wavelength.
Moreover, if the function contains oscillations with more divergent
amplitudes, e.g. $\omega^{3/2}\cos\left(\frac{2\pi}{\lambda}\omega\right)$,
one can apply the self-cancellation operator repeatedly until the
oscillations are suppressed. This general application of the self
cancellation technique we denote by 
\begin{equation}
T_{*}\left[H\left(\omega\right)\right]\equiv\left(T_{\lambda_{1}}\right)^{k_{1}}\left(T_{\lambda_{2}}\right)^{k_{2}}...\left(T_{\lambda_{n}}\right)^{k_{n}}H\left(\omega\right).\label{eq: App: Self-cancellation}
\end{equation}
For the Schwarzschild case it was argued \cite{Levi =000026 Ori - 2015 - t splitting regularization}
that there is a family of wavelengths $\lambda_{n}$ with amplitude
that decreases exponentially from $n$ to $n+1$; thus, only the first
few wavelength are important. The frequencies that correspond to the
first few wavelength $\varepsilon_{n}\equiv2\pi/\lambda_{n}$ are
\[
\varepsilon_{1}\simeq37.50\:,\;\;\;\varepsilon_{2}\simeq70.17\:,\;\;\;\varepsilon_{3}\simeq102.8\:,\;\;\;\varepsilon_{4}\simeq135.5\;.
\]

Although it is very interesting to learn the origin of the oscillations
and obtain the wavelengths by integrating the corresponding CNGs it
is not really necessary, and one can use other techniques to remove
the oscillations. This techniques are important because for some backgrounds
finding and integrating the CNGs can be a difficult task. One such
technique that we have explored is simply to use a low pass filter.
This produces very good results, though it is less efficient than
a prior knowledge of the wavelengths. Another interesting technique
that we have explored is to recursively use a Fourier transformation
of the integrand $h\left(\omega\right)$ to determine the wavelength
of the dominant oscillatory term and self-cancel it, this technique
also provided good results.

\end{document}